# Quantitative spin–orbit engineering enables subwavelength-scale programmable polarization states for versatile photonic functionalities

Shuaijie Yuan,[1,#] Xuefeng Zhang,[1,#] Jin Yang,[1,#] Min Cheng,[1] Xinyu Ma,[1] Jinhai Zou,[1] Jinyong Leng,[1] Zhongquan Nie,[1,*] Bing Lei,[1,*] and Pu Zhou[1,*]

[1]*College of Advanced Interdisciplinary Studies, National University of Defense Technology, Changsha, 410073, China*

**Correspondence: ZQ Nie, E-mail: niezhongquan1018@163.com; B Lei, E-mail: leibing_2000@nudt.edu.cn；P Zhou, E-mail: zhoupu203@163.com*

[#]*These authors contributed equally to this paper*

## ABSTRACT:

Precise and programmable control of optical polarization provides a versatile vectorial degree of freedom for light–matter interaction and optical information encoding. Existing polarization state generators (PSGs), however, suffer from poor switchability and modest fidelity at the micrometer scale, which severely constrains their practical applicability. To overcome these limitations, here we present a quantitative spin–orbit engineering that enables the dynamic synthesis of arbitrary polarization states at subwavelength dimensions. The core concept is to tailor conjugate optical vortices that independently address the orthogonal spin components of a vectorially polarized light. This spatial modulation scheme annihilates intrinsic topological charges and reshapes the spin-orbit interaction, thereby coherently constructing polarization ellipses on demand. Leveraging this mechanism, we experimentally synthesize diffraction–limited arbitrarily polarized beams spanning the entire Poincaré sphere with fidelity over 99%. As proof-of-principle demonstrations, we apply the miniaturized PSG platform to Mueller-matrix polarization microscopy, pixel-level polarization encryption, and parallel polarization-information encoding, illustrating its broad utility of programmable polarization control for subwavelength polarimetric imaging and reconfigurable multiplexed polarization encoding. Compared with state-of-the-art

PSGs, our approach uniquely integrates rapid programmability, high fidelity, and subwavelength localization in a single architecture. These multifunctional capabilities of this work position the platform as a promising route for advancing Mueller matrix polarization microscopy, multidimensional optical storage, and secure optical information processing, and beyond.

## INTRODUCTION

As a fundamental property of light, polarization offers an additional degree of freedom beyond intensity alone for optical information encoding and light–matter interaction [1,2], underpinning representative applications including polarization microscopy [3–6], polarization information encryption [7,8], polarization information encoding [9,10], and chiral sensing [11–13]. Precise generation of polarization and its dynamic manipulation are thus two long-standing goals. However, there are three key features that are both demanded and challenging to achieve in a single polarization-state generation (PSG) platform. (i) Rapid programmability for on-demand polarization-state switching across the full Poincaré sphere, allowing arbitrary states to be synthesized and switched at will; (ii) high polarization fidelity for reliable quantitative operation, which is essential for demanding applications, including high-precision Mueller-matrix polarimetry and polarization-encoded information transmission; and (iii) subwavelength localization of the polarization state, which will enable spatially confined polarization control, significantly enhancing imaging resolution and light–matter interaction strength. Therefore, there is a pressing need to develop a new PSG strategy that can simultaneously deliver rapid programmability, high fidelity, and subwavelength localization spanning the full Poincaré sphere.

Three technical routes have advanced different parts of this challenge. Rotating-retarder PSGs prepare polarization states accurately, but their mechanical switching is slow and susceptible to motion-induced errors [14]. Metasurfaces [15–18] eliminate moving elements and provide vectorial control through engineered anisotropic or phase responses; however, their functions are usually fixed at fabrication and distributed among distinct spatial channels, preventing flexible in situ switching at one focal position. SLM-based schemes remove the static limitation and synthesize dynamic vector fields [19,20]. Their common implementations, however, separate orthogonal polarization components with Ronchi gratings or Wollaston prisms and then recombine them, which increases alignment complexity and reduces optical efficiency. Diffractive neural networks and multi-pass SLM architectures [21,22] extend reconfigurable vector-field control, but require preoptimized configurations or precise inter-pass calibration. The lack of a direct physics-based mapping from programmable pupil fields to prescribed focal polarization states has therefore motivated singularity engineering. Under tight focusing, a π-phase step breaks the rotational symmetry of a cylindrically polarized beam and permits in situ switching among linear focal states only [23,24]. Helical-phase modulation instead generates localized spin angular momentum and circular-polarization components [25]. Neither approach establishes a continuous, deterministic route to arbitrary elliptical states across the full Poincaré sphere. A unified framework that combines rapid programmability, high fidelity, full-Poincaré-sphere coverage, and subwavelength focal confinement is therefore still lacking.

We bridge this gap with a quantitative spin–orbit engineering framework for dynamically programmable polarization synthesis at the subwavelength scale. A single phase-only SLM encodes conjugate vortices whose relative weights and phase independently address the left- and right-circular spin components of a radially polarized beam. Under high-NA focusing, spin–orbit coupling maps these programmable components into prescribed focal polarization states within a coaxial single-path architecture, thus constructing polarization ellipses on demand. Experimentally, we achieve full-Poincaré-sphere coverage with mean polarization errors below 1% . The proposed compact reconfigurable PSG platform further performs Mueller-matrix polarization microscopy, polarization-encoded optical encryption, and spatially multiplexed parallel encoding, delivering multiple photonic functionalities from one common programmable polarization source.

## 2. RESULTS

As illustrated in Fig. 1, a radially polarized field natively comprises orthogonal left-handed $|L\rangle$ and right-handed $|R\rangle$ circular spin bases, which are inextricably bound to the intrinsic orbital angular momentum (OAM) of the beam. In the cylindrical coordinate system, this electric field can be expressed as:

$$\mathbf{E}_{rad}(\phi) = \frac{1}{\sqrt{2}}\left(e^{i\phi}|R\rangle + e^{-i\phi}|L\rangle\right). \tag{1}$$

The spatial phases $e^{\pm i\varphi}$ represent the intrinsic OAM carried by the two circular components. Synthesizing an arbitrary polarization ellipse at the subwavelength focus requires controlling the relative complex amplitudes of the circular components that survive on axis after focusing. We achieve this by designing a customized complex amplitude mask $T(\varphi)$ composed of two conjugate vortices,

$$T(\phi) = \sqrt{w}e^{-i\phi} + \sqrt{1-w}e^{i\delta}e^{i\phi}. \tag{2}$$

Within this modulation scheme, the weighting factor $w$ in the mask directly governs the energy distribution between the two conjugate vortices, while $\delta$ dictates their relative phase retardation. The modulated field $\mathbf{E}_{mod}$ thus comprises four distinct spin-orbit coupled states,

$$\mathbf{E}_{mod}(\phi) = \frac{1}{\sqrt{2}}\left[\sqrt{w}|R\rangle + \sqrt{1-w}e^{i\delta}|L\rangle + \sqrt{1-w}e^{i\delta}e^{i2\phi}|R\rangle + \sqrt{w}e^{-i2\phi}|L\rangle\right]. \tag{3}$$

These generated states are then tightly focused by a high-numerical-aperture (NA) objective, and their focal distribution is governed by the Richards-Wolf vectorial diffraction integral [26-30],

$$\mathbf{E}(r,\phi) = -\frac{ikf}{2\pi}\int_0^{\alpha_{max}}\int_0^{2\pi} P(\theta)\sqrt{\cos\theta}\mathbf{R}(\theta,\varphi)\mathbf{E}_{\mathrm{mod}}(\varphi)e^{ikr\sin\theta\cos(\varphi-\phi)}\sin\theta d\varphi d\theta. \tag{4}$$

Here, $k$ is the wavenumber, $f$ is the focal length, $\alpha_{max}$ is the maximum semi-aperture angle, and $(r, \phi)$ are the cylindrical coordinates. This tight-focusing mechanism inherently drives a robust vectorial spatial filtering process. Because the generated high-order spatial modes and longitudinal depolarization fields carry intrinsic central phase singularities, their energies are physically expelled into peripheral annular regions. Consequently, only the desired zero-order transverse components survive at the exact focal center ( $r = 0$ ). With their topological charges annihilated, these surviving components coherently superpose to form an analytical field,

$$\mathbf{E}_f(r=0) \propto \sqrt{1-w}e^{i\delta}|L\rangle + \sqrt{w}|R\rangle. \tag{5}$$

This analytical on-axis field represents an arbitrary pure transverse polarization state on the Poincaré sphere. We use $|L\rangle = (\mathbf{e}_x + i\mathbf{e}_y)/\sqrt{2}$, $|R\rangle = (\mathbf{e}_x - i\mathbf{e}_y)/\sqrt{2}$, and $S_3 = I_R - I_L$. By matching this focal distribution to a target polarization ellipse characterized by the azimuth angle $\chi$ and the ellipticity angle $\eta$, we directly obtain the exact mapping equations:

$$\mathbf{E}(\chi,\eta) = \cos\left(\frac{\pi}{4} - \eta\right)e^{i\chi}|R\rangle + \sin\left(\frac{\pi}{4} - \eta\right)e^{-i\chi}|L\rangle. \tag{6}$$

$$w = \frac{1+\sin(2\eta)}{2}, \delta = -2\chi. \tag{7}$$

These unified relations establish a rigorous quantitative spin-orbit coupling framework. Continuous tuning of the modulation weights and relative phases enable deterministic navigation of the synthesized subwavelength polarization states across the entire Poincaré sphere. Specifically, the rightmost table in Fig. 1 summarizes these modulation parameters. For pure linear polarizations, the weighting factor is strictly set to 0.5, while the relative phase determines the spatial orientation. Conversely, for pure circular polarizations, this factor is either 1 or 0. For generalized elliptical polarizations, both parameters cooperatively dictate the polarization ellipse, as depicted by Eq. (7). The complete derivation is provided in Section S1 of the Supplementary Information.

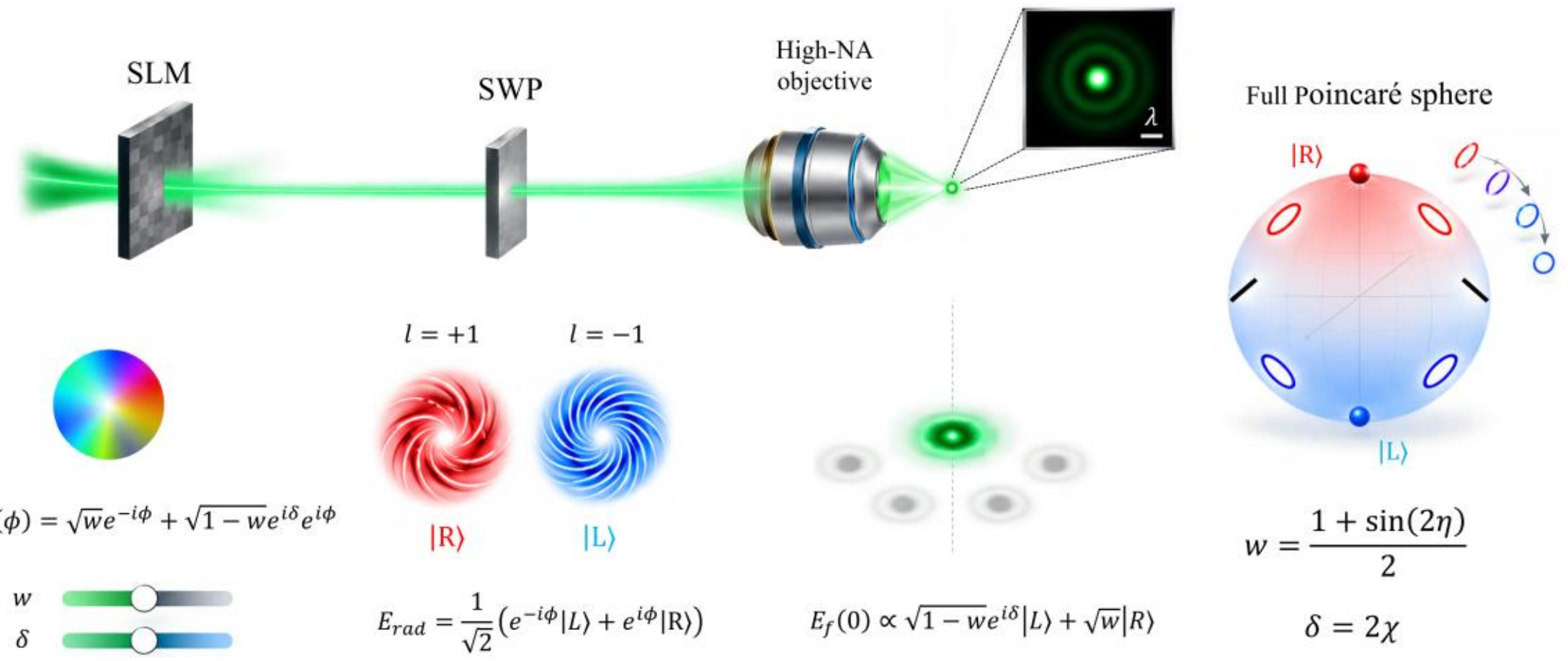

**Fig. 1.** Conceptual framework for deterministic synthesis of subwavelength polarization states through quantitative spin-orbit coupling.

According to our theoretical framework, controlling the ellipse morphology requires the superposition of positive and negative conjugate vortices. A phase-only mask cannot directly set the amplitude in its own plane. We therefore use interference and spatial filtering to encode the required complex amplitude. We implement this using the double-phase hologram superpixel encoding technique, illustrated in Fig. 2a. In this algorithm, four adjacent SLM pixels are grouped into one macro pixel, where diagonal subpixels share identical phase values. By assigning specific phase values $\phi_1$ and $\phi_2$ via an arccosine function, the Euler expansion of the pure-phase terms yields an effective amplitude envelope. Far field coherent superposition spatially averages these discrete sub pixels to reconstruct the desired complex amplitude [31,32]:

$$\widetilde{U}_{target} = \frac{1}{2}\left(\widetilde{U}_1 + \widetilde{U}_2\right) = Ae^{i\Phi}. \tag{8}$$

This configuration enables coherent interference at the focal plane to independently modulate both amplitude and phase. Substituting the weight $w$ and relative phase $\delta$ for the required azimuth angle $\chi$ and ellipticity angles $\eta$ into this algorithm yields our analytical target parameters as follows:

$$A(\varphi) = \sqrt{1 + 2\sqrt{w(1-w)}\cos(2\varphi + \delta)}. \tag{9}$$

$$\phi(\varphi) = \arctan\left[\frac{-\sqrt{w}\sin(\varphi) + \sqrt{1-w}\sin(\varphi + \delta)}{\sqrt{w}\cos(\varphi) + \sqrt{1-w}\cos(\varphi + \delta)}\right]. \tag{10}$$

We then calculate the exact phase distribution required for each sub-pixel $\phi_1 = \phi + \arccos A$ and $\phi_2 = \phi - \arccos A$. Using this wavefront engineering, we construct the dynamic polarization state generation optical path shown in Fig. 2b. A 532 nm continuous-wave laser passes through a linear polarizer to obtain highly pure horizontal polarized light. This beam illuminates the SLM loaded with the double-phase hologram. The modulated vector beam enters the 4f lens system, where the first lens performs the spatial Fourier transform. A low-pass filter located at the Fourier plane blocks high-frequency diffraction orders. The second lens then executes an inverse Fourier transform, reconstructing the target complex-amplitude field.

Following spatial wavefront shaping, the beam passes through the S-wave plate (SWP) with a topological charge of 0.5. This geometric-phase element converts the complex-amplitude field into a structured field carrying customized local spin components. This coaxial propagation path relaxes optical alignment constraints by bypassing previous vector beam generators that require rigorous alignment for separate interferometric arms. The beam is then focused by an objective lens with NA = 0.95 under nonparaxial conditions. Strong spin-orbit coupling compresses the focal spot to the subwavelength scale, providing ultrahigh spatial resolution, while simultaneously pushing unwanted high-order modes into the peripheral region. These tight-focusing conditions ensure that the target polarization state is synthesized precisely at the focal center. Insets in Fig. 2b illustrate a representative target: an elliptical polarization with an azimuth angle of 60° and an ellipticity angle of 30°. We display required field distributions at key optical planes to track its physical evolution. The transverse intensity spot morphology at the focal plane depends strongly on the polarization state while deviates completely from conventional Airy disks. This unique spatial distortion arises from cross-polarization crosstalk and vectorial field redistribution. To quantitatively evaluate this focal-plane field, we employ a classical rotating quarter-wave-plate polarimeter module. The wave plate is rotated in 22.5° increments and CMOS camera captures eight raw intensity images to perform Stokes polarization imaging. Simulation results show excellent agreement with predesigned modulation polarization strictly within the full width at half maximum intensity region.

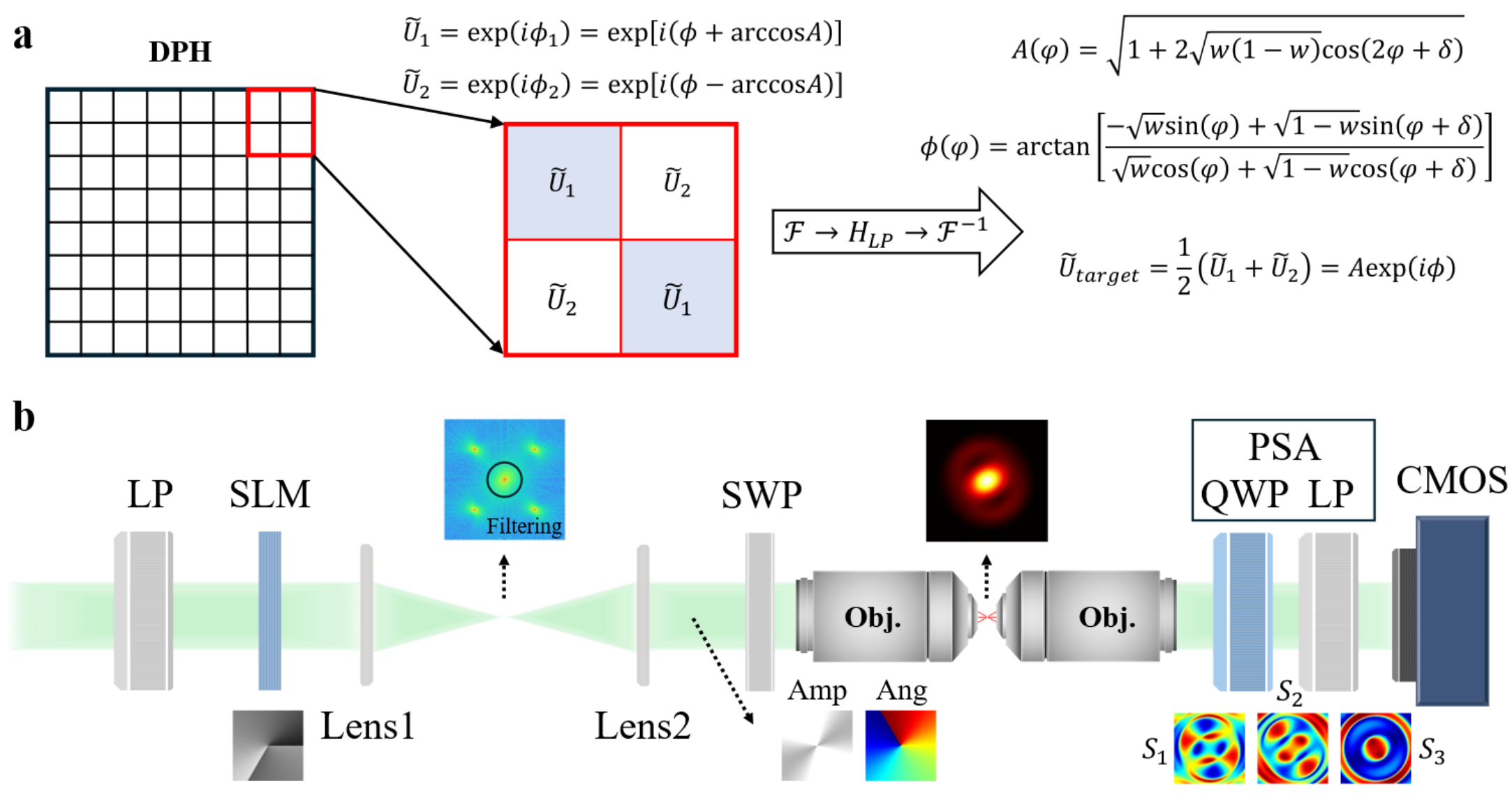

**Fig. 2.** Complex-amplitude encoding. (a) Double-phase hologram (DPH) superpixel encoding with spatial low-pass filtering for complex-amplitude reconstruction. (b) Optical configuration for spatial modulation, Fourier filtering, SWP transformation, high-NA focusing, and Stokes polarimetry.

We next select eight representative polarization states for detailed morphological analysis to comprehensively validate our approach. These target states include horizontal linear, vertical linear, right-handed circular (ellipticity angle of 45°), left-handed circular (ellipticity angle of −45°), and four arbitrary elliptical polarization states. Figure S2 of the Supplementary Information details the theoretical generation process and the expected focal fields for these targets. From left to right, the panels first display the required incident amplitude and phase distributions, denoted as $\mathrm{Amp}(E_i)$ and $\mathrm{Ang}(E_i)$, at the pupil plane before the objective. These specific complex pupil fields are synthesized using the adjacent pure-phase masks loaded onto the SLM through complex-amplitude encoding. The subsequent panels exhibit the simulated transverse focal intensity $I_r$, deliberately excluding the longitudinal field component $I_z$ to focus strictly on the transverse plane. The rightmost columns present the theoretical Stokes parameters $S_1$, $S_2$, and $S_3$, respectively, establishing a baseline for polarization imaging. These ideal simulated profiles allow us to accurately isolate and evaluate the inevitable cross-polarization coupling effects that arise in practical tight focusing optical systems.

Figure 3 presents the corresponding experimental results. These panels display the measured transverse focal intensities, the calculated Stokes parameter distributions, and the reconstructed polarization ellipses. Additional experimental data for the remaining elliptical states are provided in Fig. S7 of the Supplementary Information. In tight-focusing systems, the nonparaxial condition inherently induces spatial depolarization and cross-polarization coupling at the focal periphery. Therefore, to accurately compute the final experimental Stokes parameters, our spatial averaging is confined to the central region where the intensity exceeds 80% of the maximum. Applying this criterion, the measured experimental Stokes vectors $[S_1, S_2, S_3]^T$ for the four fundamental representative states are $[0.9694, -0.0028, -0.0517]^T$ for horizontal linear, $[-0.9548, 0.0234, 0.0179]^T$ for vertical linear, $[0.0455, -0.0038, 0.9728]^T$ for right-handed circular, and $[-0.0269, -0.0124, -0.9843]^T$ for left-handed circular. The measurements yield exceptionally low errors of 2.84%, 2.88%, 2.55%, and 1.83%, respectively. The remaining four elliptical states also show agreement with the prescribed polarization parameters. These measurements support programmable focal polarization synthesis within the stated sampling region.

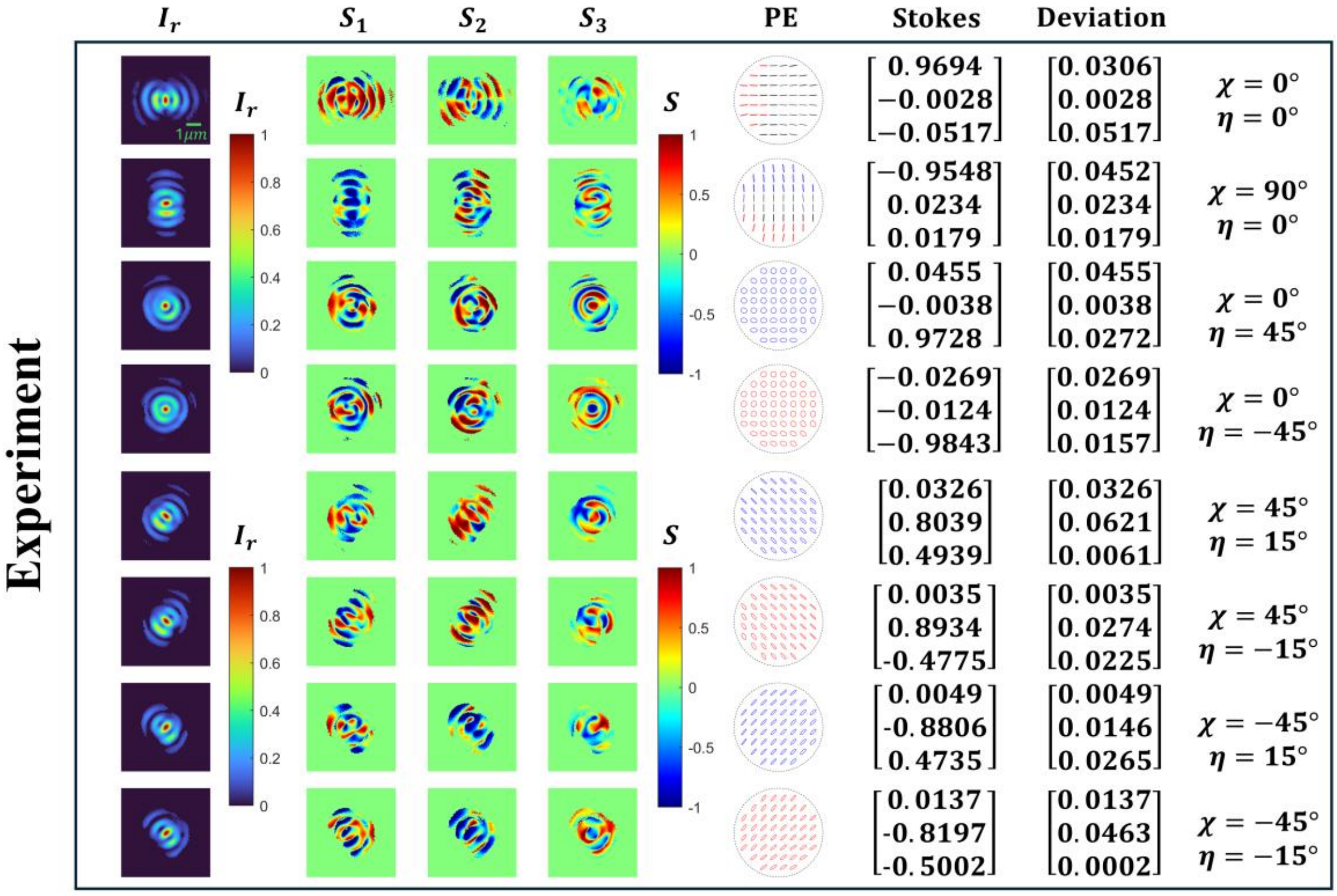

**Fig. 3.** Experimental focal Stokes imaging for representative polarization states. The transverse intensity, measured Stokes-parameter distributions, and reconstructed polarization ellipses (PE) are shown. Stokes vectors are spatially averaged within the contour at 80% of the peak intensity.

To assess the state-generation accuracy of our quantitative spin-orbit coupling framework, we extend our validation to 22 sampled states distributed across the entire Poincaré sphere. Figure 4 presents a quantitative characterization of the measured Stokes parameters obtained with two distinct data extraction strategies. The first approach extracts data exclusively from the exact focal center. Under this condition, Fig. 4a illustrates the generation error distributions, and Fig. 4b presents the corresponding Poincaré sphere mapping, where hollow triangles indicate the theoretical target values while colored solid circles represent the experimental measurements. The color map indicates the absolute generation error, defined as the arithmetic mean of the absolute deviations across the three Stokes parameters. The bar chart in Fig. 4c provides a clearer comparative analysis of the overall parameter deviations. At the exact focal center, the individual errors for the three Stokes parameters are 0.74%, 0.66%, and 1.01% respectively, yielding a global average error of only 0.8%. This experimentally measured value at the micro/nanoscale regime approaches the physical measurement limit of current macroscopic polarization detection systems.

Expanding the validation boundary, the second approach calculates the spatial average within the region where the intensity exceeds 80% of the maximum. Figure 4d displays the corresponding error distributions, extracted with its Poincaré sphere mapping and bar chart detailed in Figs. 4e and 4f. Under this broader spatial averaging condition, the individual deviations elevated to 2.71%, 2.46%, and 3.74%, resulting in a global average error of 2.97%. This localized purity degradation is a physically inevitable consequence of the inherent cross-polarization coupling effect under high-numerical-aperture tight focusing.

Under both extraction conditions, the generation error for $S_3$ is consistently larger than that for the two linear Stokes components. This discrepancy arises because $S_3$ governs the circular polarization properties, yet its generation faces practical conversion efficiency limits inherent to spin-orbit coupling process. In addition, slight physical misalignments between the center of the complex-amplitude distribution generated by the SLM and the exact center of the waveplate disproportionately degrades the chiral purity. Despite these practical sensitivities, the resulting error distributions vary smoothly across the sampled states. This robust performance confirms the universal applicability and exceptional accuracy of our dynamic synthesis paradigm across the entire Poincaré sphere.

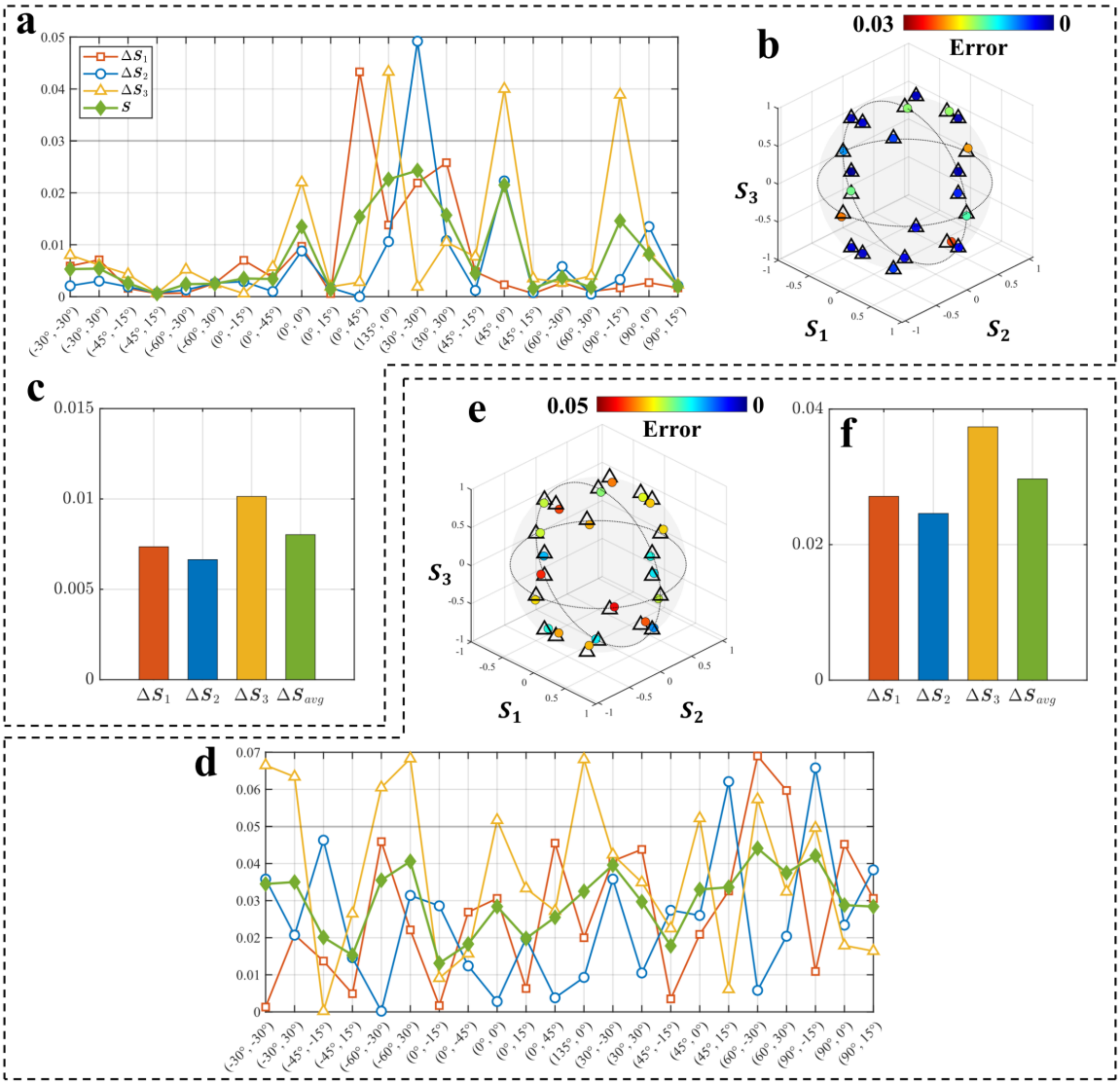

**Fig. 4.** Quantitative validation of generated polarization states. (a,d) Absolute deviations of the three normalized Stokes components and their per-state mean absolute error for 22 states, evaluated at the focal center and within the 80% peak-intensity contour, respectively. (b,e) Corresponding Poincaré-sphere maps; open triangles indicate target states, filled circles indicate measurements, and color encodes the per-state mean absolute error. (c,f) Component-wise mean absolute errors across the 22 states and their overall mean.

The subwavelength polarization control established above naturally supports Mueller matrix microscopy, which benefits directly from rapid reconfigurability, high fidelity, high-resolution spatial localization, and full Poincaré sphere accessibility. We therefore applied this miniaturized PSG platform to a laboratory-prepared R5011-doped E7 chiral liquid crystal (see Supplementary Section S4 for the preparation procedure and structural evolution). R5011 introduces a chiral twist into the orientational order of nematic E7 host and drives the mixture into a chiral nematic phase characterized by a helically varying director field and strongly polarization-dependent optical response. After centrifugation, an aliquot of the dispersion was deposited onto a clean glass slide, dried, and imaged under an optical microscope. The optical micrograph in Fig. 5a shows isolated annular and droplet-like features distributed across the field of view. The SEM image in Fig. 5b reveals a representative liquid-crystal aggregate with characteristic projected dimensions of 0.6–0.9 μm, comparable to the operating wavelength of λ = 532nm.

Figure 5d shows that the normalized Mueller matrix of this aggregate exhibits pronounced spatial heterogeneity in its polarization response. The most distinctive feature appears in $m_{03}$, where a negative central region is surrounded by positive annular and arc-shaped features. Since $m_{03}$ describes the differential response to right- and left-circular input states, this contrast indicates spatially varying circular diattenuation within the chiral liquid crystal. The $m_{30}$ element, by contrast, reports circular polarizance. Substantial off-diagonal content in $m_{13}$, $m_{23}$, $m_{31}$, and $m_{32}$ points to coupling between linear and circular Stokes components, while spatial variations in $m_{01}$, $m_{02}$, $m_{10}$, and $m_{20}$ reveal linear diattenuation and polarizance. Taken together, these elements resolve the coexistence of circular selectivity, linear anisotropy, and linear-to-circular polarization conversion at the wavelength scale, information that is absent from conventional intensity images.

For comparison, Fig. 5e presents the Mueller matrix imaging of a 1 μm polystyrene microsphere (Fig. 5c). Its normalized Mueller matrix remains close to the identity matrix, with diagonal elements near unity and off-diagonal elements close to zero, consistent with a weakly polarizing, near-isotropic scatterer. The performance of our platform can be compared with that of state-of-the-art Mueller matrix microscopes from two perspectives. First, full-Poincaré-sphere programmability allows the four probing Stokes states to be placed precisely at the vertices of a regular tetrahedron on the Poincaré sphere. This minimizes the condition number of the state-generation matrix and therefore improves the noise robustness of matrix reconstruction. Second, rapid in situ switching at a fixed focal position eliminates mechanical rotation and the associated beam drift. Metasurface-based alternatives, in contrast, typically encode different probing states in separate spatial channels, which can introduce pixel-level misregistration.

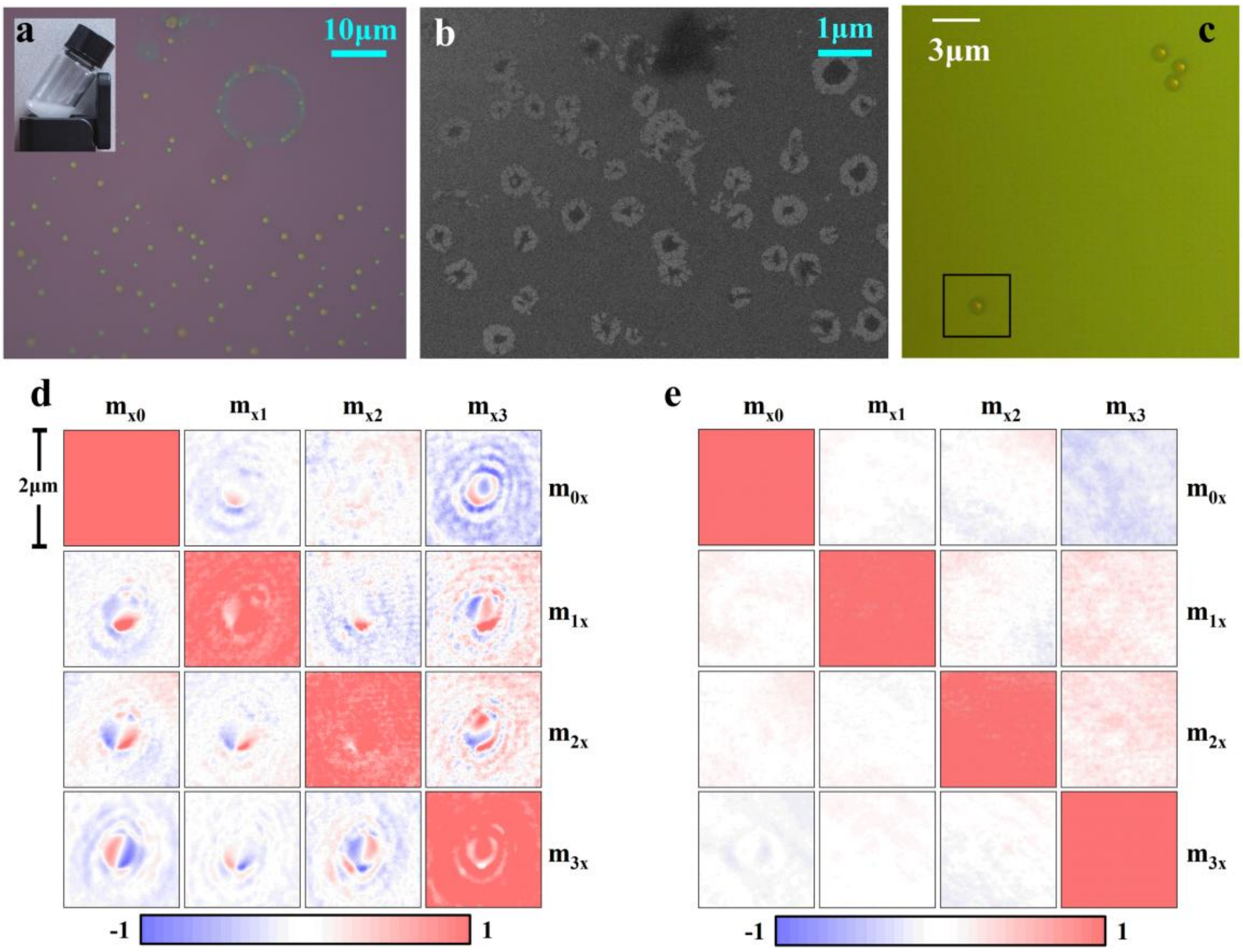

**Fig. 5. Mueller matrix polarization imaging of wavelength-scale chiral liquid-crystal microstructures and polystyrene microspheres.** (a) Optical microscopy image of the R5011-doped E7 liquid-crystal. (b) Scanning electron microscopy image of a representative liquid-crystal aggregate (magnification factor: 10kx). (c) Optical microscopy image of 1 μm polystyrene microspheres. (d) Normalized Mueller matrix images of the chiral liquid-crystal microstructure in a. (e) Mueller matrix images of the polystyrene microspheres in c.

To verify the dynamic programmability of the polarization-state source and its use in pixel-level polarization encryption, we constructed the free-space transmission system shown in Fig. 6a. A prescribed polarization state is generated at the focal plane and collected by two high-NA objectives. The output beam is subsequently expanded and collimated by a two-lens telescope and then propagates 1m in free space. The transmitted polarization state is measured with a commercial polarimeter (Thorlabs PAX1000). We selected four linear polarization states at 0°, 45°, 90°, −45° as the physical alphabet and assigned each to a distinct color symbol. For the $n$-th pixel, the input symbol $q_n \in \{0, 1, 2, 3\}$ is encrypted by modular addition with a pixel-dependent secret key $k_n \in \{0, 1, 2, 3\}$ according to $p_n = (q_n + k_n) \bmod 4$, which determines the transmitted polarization state. The original symbol is recovered by the inverse operation $q_n = (p_n - k_n) \bmod 4$. With 4,096 pixels and 4 key values per pixel, the key space contains $4^{4096}$ distinct keys, rendering brute-force attacks computationally infeasible. This four-state alphabet demonstrates the principle of polarization-domain optical encryption. A communication implementation could instead combine independently generated, nonreused key streams with noise-adapted polarization constellations, including elliptical states distributed over the Poincaré sphere. These choices can further improve symbol discrimination and coding flexibility.

During transmission, the SLM switches to a new polarization state every 500 ms, while the polarimeter records a normalized Stokes vector every 50 ms. To suppress switching transients, the first and last samples of each 500 ms interval were discarded, and the remaining central samples were averaged for state identification. Figures 6b1 and 6b2 show the complete Stokes sequences measured over all 4096 pixels of the two test images. As expected for the four linear states, $S_3/S_0$ remains close to zero throughout the transmission, whereas $S_1/S_0$ and $S_2/S_0$ repeatedly take their prescribed positive, negative and near-zero levels. The persistence of these distinct levels over the entire payload demonstrates stable, state-by-state polarization switching during continuous free-space transmission.

Figures. 6c1 and 6c2 visualize the state separation. All 4096 received states cluster tightly around the four prescribed positions on the equator of the Poincaré sphere. The compact intrastate distributions and clear separation between neighboring clusters indicate that repeated switching introduces only limited fluctuations while retaining sufficient contrast for reliable classification. The residual broadening relative to the static measurements is consistent with the finite settling time of the SLM, the response and integration time of the polarimeter, residual synchronization mismatch, and small polarization fluctuations accumulated during propagation. These effects broaden the measured clusters but do not obscure the four encoded states.

Figures 6d1 and 6d2 show the target images and decoded results, thereby verifying the complete encryption–transmission–decryption chain. With the correct key, both the Rubik's cube and the university emblem are recovered without symbol errors. Without the inverse key, the recovered images are indistinguishable from random noise, with symbol error rates of 3076/4096 ≈ 75.1% and 3091/4096 ≈ 75.5%, respectively. This agreement confirms that the transmitted information remains secure against unauthorized decoding. In the present experiment, each transmitted state is verified by sequential polarimeter readout; the demonstrated symbol rate is therefore governed primarily by the 500 ms state dwell time and serial Stokes acquisition. This proof-of-principle rate therefore does not represent the throughput limit of the encoding strategy. Faster spatial light modulators can shorten the switching interval, while parallel detection can replace serial polarimetric readout and substantially increase the receiver throughput. Moreover, extending the four equatorial states to arbitrary elliptical states across the full Poincaré sphere would enlarge both the polarization alphabet and the available key-dependent mappings without changing the source architecture.

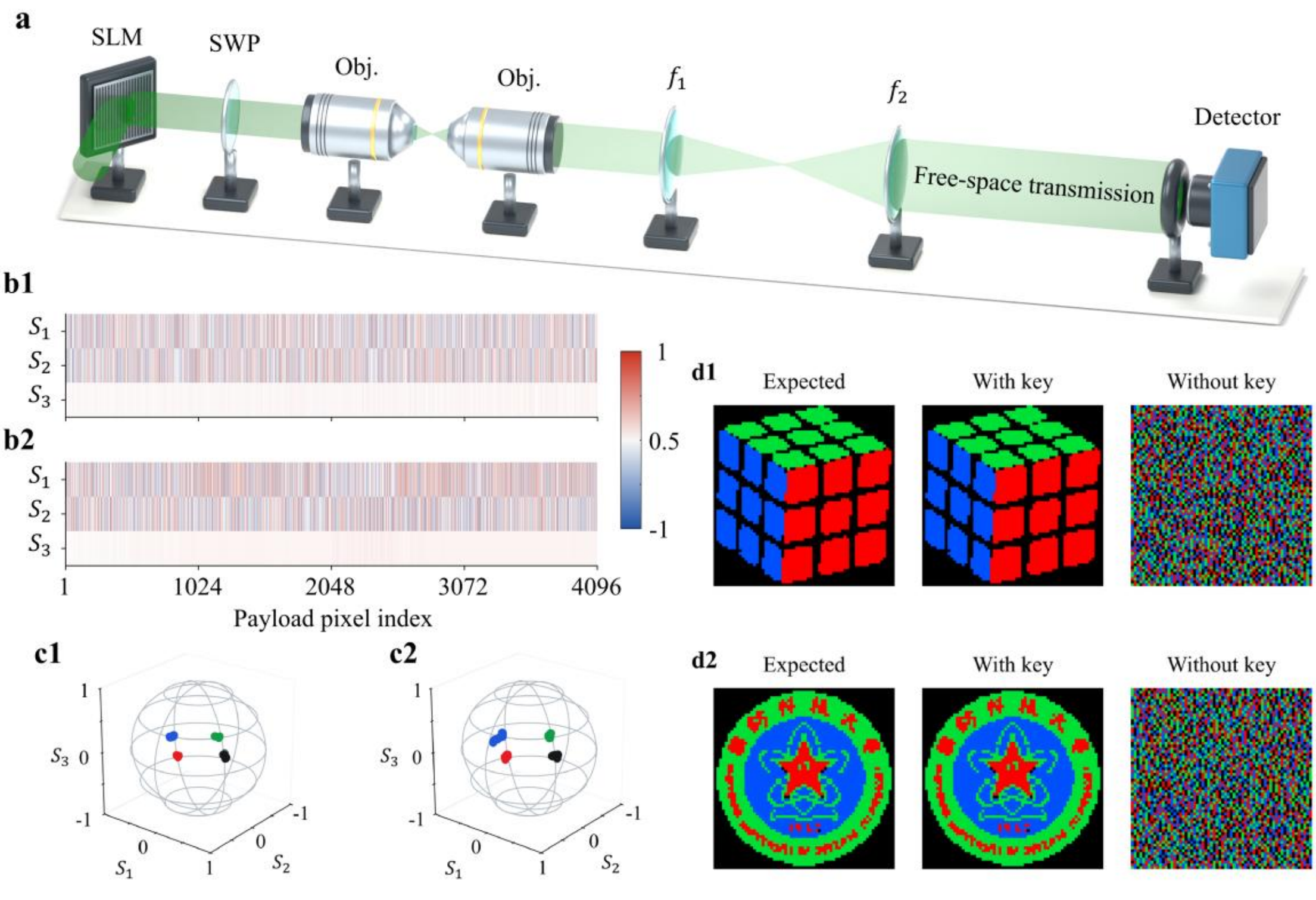

**Fig. 6.** Experimental demonstration of pixel-level polarization encryption for free-space transmission. (a) Free-space transmission and polarization detection configuration. (b1) and (b2) Measured Stokes sequences of two encrypted 64 × 64 images over 4096 transmitted pixels. (c1) and (c2) Corresponding measured polarization-state distributions on the Poincaré sphere. (d1) and (d2) Target images (left) and decoded results obtained with the correct key (middle) and without it (right).

The sequential Stokes readout provides direct verification of the transmitted polarization states, but it also limits the acquisition throughput. We therefore further explored camera-based intensity detection as a more scalable readout scheme. Here, two linear polarization states are assigned to the binary symbols black and white. Two spatial channels are generated simultaneously, giving four possible joint states corresponding to the binary combinations 00, 01, 10, and 11. Each combination produces a distinct spatial intensity pattern at the receiver. As shown in Fig. 7a, the measured distributions closely reproduce the corresponding simulations (insets) and remain clearly distinguishable. The encoded state can therefore be identified directly from its spatial intensity signature without reconstructing the full Stokes vector. Together, these four binary combinations define four distinct polarization-encoded intensity states.

This spatially resolved response was then used for parallel pixel-level encoding and encrypted transmission. According to the encrypted data stream, the corresponding phase masks are loaded sequentially onto the SLM, as illustrated in Fig. 7b1. The camera records the two spatial channels simultaneously and produces the received intensity-pattern sequence shown in Fig. 7b2. In this way, a single camera exposure reads the complete joint state, rather than interrogating each polarization channel separately. The clear correspondence between the loaded and detected sequences confirms stable discrimination of the four binary combinations during repeated switching. The same principle is not restricted to two channels. Larger polarization-state arrays, including two-dimensional configurations, can be generated for higher-degree parallel encoding, as further demonstrated in Supplementary Section S5.

The complete encoding-transmission-decoding chain was finally tested using two QR codes carrying the messages "hello NUDT" and "GOOD". With the correct key, both decoded QR codes faithfully reproduce the target patterns, as shown in Fig. 7c1 and Fig. 7c2. Removing the corresponding key destroys the pixel correspondence and produces random-like binary patterns, in which the encoded information is no longer recognizable. Importantly, the camera-based scheme converts the sequential polarization readout demonstrated above into simultaneous spatial detection, directly exploiting the multiplexing capability of the programmable polarization source.

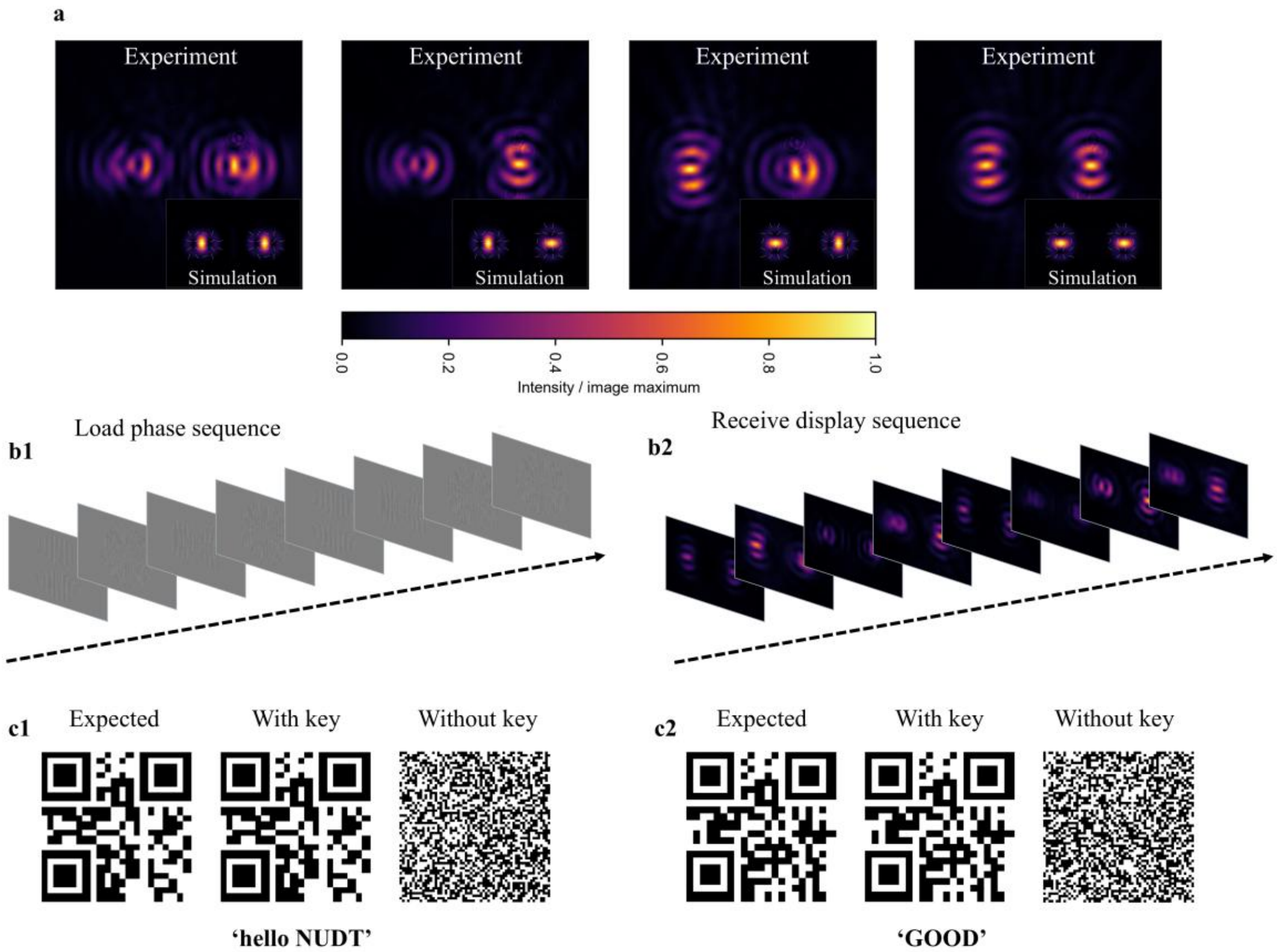

**Fig. 7.** Experimental intensity patterns of the four joint binary states, ordered as 00, 01, 10 and 11 from left to right, with the corresponding simulations shown in the insets. (b1) Sequentially loaded phase masks for pixel-wise polarization encoding. (b2) Corresponding received intensity-pattern sequence recorded by the camera. (c1) and (c2) Target QR codes (left column), correctly decoded results (middle column) and results obtained without the correct key (right column) for the messages "hello NUDT" and "GOOD", respectively.

## 3. CONCLUSION AND DISCUSSION

In this work, we establish a deterministic mapping from programmable conjugate-vortex parameters to focal polarization under strong spin–orbit coupling. By independently tailoring the conjugate optical vortices that address the orthogonal spin components of a vectorially polarized beam, we coherently construct arbitrary polarization ellipses on demand and reconstruct target states across the full Poincaré sphere at the subwavelength scale, with a mean experimental error below 1% over the tested state ensemble. The proposed PSG platform uniquely integrates high-fidelity state synthesis, subwavelength localization, dynamic reconfigurability and spatial multiplexing within a single architecture.

We demonstrate the versatility of this miniaturized PSG approach through Mueller-matrix polarization microscopy and pixel-level polarization encryption, and parallel polarization-information encoding. First, the subwavelength polarimetric imaging resolves spatially heterogeneous circular diattenuation, linear anisotropy and linear–circular coupling in a wavelength-scale chiral liquid crystal that are inaccessible to conventional intensity imaging. Second, the optical polarization encryption achieves error-free recovery with the correct key and noise-like outputs without it, with symbol error rates consistent with the theoretical expectation for a uniformly random incorrect key. It should be noted that the present SLM has a nominal maximum refresh rate of 60 Hz, corresponding to a frame interval of 16.7 ms. The longer dwell used in the transmission experiments accommodates optical settling, synchronization and polarimeter acquisition. The approximately 500 ms SLM update and serial Stokes acquisition nevertheless set the present throughput, so the demonstrated rate is a proof-of-principle bound rather than a fundamental limit. Faster phase-only SLMs, or DMD-based holography with appropriate complex-field encoding, could increase pattern-update rates. Third, intensity-based parallel polarization encoding offers a scalable alternative, but its performance hinges on the trade-off among channel cross-talk, photon budget and calibration stability. Likewise, the current physical-layer encoding demonstrates key-dependent concealment, not unconditional security.

Looking forward, replacing serial Stokes acquisition with simultaneous spatial-sampling detection could transform the measurement chain into single-exposure polarimetry. Scaling a single focal state to a regular array of independently addressable subwavelength polarization states would enable full-field parallel polarimetry and higher-dimensional encoding. Integrating forward error correction, expanding the alphabet to arbitrary elliptical states across the full Poincaré sphere, and systematically characterizing switching repeatability, long-term drift and inter-channel crosstalk are necessary steps before the platform can be fairly benchmarked against high-speed commercial PSGs. Taken together, these advances may elevate programmable focal polarization from a PSG technique to a reconfigurable photonic resource for for Mueller matrix polarization microscopy, multidimensional optical storage and vectorial optical information processing.

**Funding.** National Natural Science Foundation of China (Grant No. 12574333, 11974258, 62475285, 61975235); High-Level Scientific and Technological Innovation Talent Program (Grant No. 202601-YJRC-ZZ), Innovation Science Fund of National University of Defense Technology (Grant No. 25-28); Hunan Graduate Innovation Project (Grant No. CX20240112).

**Disclosures.** The authors have no conflicts to disclose.

**Author Contributions.** S. Yuan and Z. Nie conceived this idea and conducted theoretical research. S. Yuan conducted the experiment and analyzed the data with assistance from X. Zhang and J. Yang. Meanwhile, all the authors have contributed to the writing of this paper. Z. Nie, P. Zhou and B. Lei supervised this project.

**Data Availability.** The data that support the plots within this paper and other findings of this study are available within this article and its Supplementary Information file are also available from the corresponding author upon request.

# REFERENCES

1. J. P. Balthasar Mueller, N. A. Rubin, R. C. Devlin, *et al.*, "Metasurface polarization optics: independent phase control of arbitrary orthogonal states of polarization," Phys. Rev. Lett. 118, 113901 (2017).
2. A. Forbes, M. de Oliveira, and M. R. Dennis, "Structured light," Nat. Photonics 15, 253–262 (2021).
3. S. Y. Lu and R. A. Chipman, "Interpretation of Mueller matrices based on polar decomposition," J. Opt. Soc. Am. A 13, 1106–1113 (1996).
4. C. He, H. He, J. Chang, *et al.*, "Polarisation optics for biomedical and clinical applications: a review," Light Sci. Appl. 10, 194 (2021).
5. E. Du, H. He, N. Zeng, *et al.*, "Mueller matrix polarimetry for differentiating characteristic features of cancerous tissues," J. Biomed. Opt. 19, 076013 (2014).
6. S. Liu, W. Du, X. Chen, *et al.*, "Mueller matrix imaging ellipsometry for nanostructure metrology," Opt. Express 23, 17316–17329 (2015).
7. X. Li, T.-H. Lan, C.-H. Tien, et al., "Three-dimensional orientation-unlimited polarization encryption by a single optically configured vectorial beam," Nat. Commun. 3, 998 (2012).
8. X. Ouyang, Y. Xu, M. Xian, et al., "Synthetic helical dichroism for six-dimensional optical orbital angular momentum multiplexing," Nat. Photonics 15, 901–907 (2021).
9. Z. Q. Nie, H. Lin, X. F. Liu, *et al.*, "Three-dimensional super-resolution longitudinal magnetization

spot arrays," Light Sci. Appl. 6, e17032 (2017).

10. P. Zijlstra, J. W. M. Chon, and M. Gu, "Five-dimensional optical recording mediated by surface plasmons in gold nanorods," Nature 459, 410–413 (2009).
11. Y. Tang and A. E. Cohen, "Optical chirality and its interaction with matter," Phys. Rev. Lett. 104, 163901 (2010).
12. L. A. Warning, A. R. Miandashti, L. A. McCarthy, *et al.*, "Nanophotonic approaches for chirality sensing," ACS Nano 15, 15538–15566 (2021).
13. Y. Zhao, A. N. Askarpour, L. Sun, *et al.*, "Chirality detection of enantiomers using twisted optical metamaterials," Nat. Commun. 8, 14180 (2017).
14. M. H. Smith, "Optimization of a dual-rotating-retarder Mueller matrix polarimeter," Appl. Opt. 41, 2488–2493 (2002).
15. P. C. Wu, W.-Y. Tsai, W. T. Chen, et al., "Versatile polarization generation with an aluminum plasmonic metasurface," Nano Lett. 17, 445–452 (2017).
16. F. Ding, B. Chang, Q. Wei, et al., "Versatile polarization generation and manipulation using dielectric metasurfaces," Laser Photonics Rev. 14, 2000116 (2020).
17. Y. Pan, H. Liu, Z. Cheng, *et al.*, "Supercell-based metasurfaces for arbitrary polarization beam splitting: physics-informed U-Net design with high extinction ratio," Opt. Express 33, 39960–39976 (2025).
18. Y. Gao, Q. Chen, and Y. Ma, "Non-Hermitian geometric phase metasurfaces for linearly polarized light," Optica 13, 984–990 (2026).
19. H. Chen, J. Hao, B. F. Zhang, *et al.*, "Generation of vector beam with space-variant distribution of both polarization and phase," Opt. Lett. 36, 3179–3181 (2011).
20. Z. Chen, T. Zeng, B. Qian, *et al.*, "Complete shaping of optical vector beams," Opt. Express 23, 17701–17710 (2015).
21. H. Hu, G. Wang, L. Wang, *et al.*, "Multi-mode vector vortex beams generation enabled by a rotatable D2NN," Opt. Express 34, 743–753 (2026).
22. F. Kenny, D. Lara, O. G. Rodríguez-Herrera, *et al.*, "Complete polarization and phase control for focus-shaping in high-NA microscopy," Opt. Express 20, 14015–14029 (2012).
23. H. Ren, X. Li, and M. Gu, "Polarization-multiplexed multifocal arrays by a π-phase-step-modulated azimuthally polarized beam," Opt. Lett. 39, 6771–6774 (2014).
24. L. Zhu, M. Sun, D. Zhang, *et al.*, "Multifocal array with controllable polarization in each focal spot," Opt. Express 23, 24688–24698 (2015).
25. P. Yu, Q. Zhao, X. Hu, *et al.*, "Orbit-induced localized spin angular momentum in the tight focusing of linearly polarized vortex beams," Opt. Lett. 43, 5677–5680 (2018).
26. K. Y. Bliokh, F. J. Rodríguez-Fortuño, F. Nori, *et al.*, "Spin–orbit interactions of light," Nat. Photonics 9, 796–808 (2015).
27. J. Chen, L. Yu, C. Wan, *et al.*, "Spin–orbit coupling within tightly focused circularly polarized spatiotemporal vortex wavepacket," ACS Photonics 9, 793–799 (2022).
28. O. G. Rodríguez-Herrera, D. Lara, K. Y. Bliokh, *et al.*, "Optical nanoprobing via spin-orbit interaction of light," Phys. Rev. Lett. 104, 253601 (2010).
29. K. S. Youngworth and T. G. Brown, "Focusing of high numerical aperture cylindrical-vector beams," Opt. Express 7, 77–87 (2000).
30. B. Richards and E. Wolf, "Electromagnetic diffraction in optical systems, II. Structure of the image field in an aplanatic system," Proc. R. Soc. Lond. A 253, 358–379 (1959).

31. V. Arrizón and D. Sánchez-De-la-Llave, “Double-phase holograms implemented with phase-only spatial light modulators: performance evaluation and improvement,” Appl. Opt. 41, 3436–3447 (2002).
32. V. Arrizón, “Complex modulation with a twisted-nematic liquid-crystal spatial light modulator: double-pixel approach,” Opt. Lett. 28, 1359–1361 (2003).